\documentclass[AMA,STIX1COL,svgnames]{WileyNJD-v2}

\usepackage[framemethod=TikZ]{mdframed}
\usepackage{amsmath,fixmath,listing,color,mathastext,graphicx,wrapfig}
\usepackage{tikz}
\usepackage{array}
\usetikzlibrary{positioning}
\usetikzlibrary[shapes.misc]
\usetikzlibrary{arrows,automata}
\usepackage{todonotes}
\usepackage{threeparttable}
\usepackage{hyperref}
\hypersetup{colorlinks=true, linkcolor=blue, filecolor=blue, urlcolor=blue, citecolor=black}
\usepackage{listings}
\usepackage{xcolor}
\usepackage{xkeyval} 
\input{listings-stata.tex}
\PassOptionsToPackage{hyphens}{url}
\usepackage{url}

\usepackage{arydshln}
\usepackage{tabularray,lipsum,xcolor}

\usepackage{csquotes}
\usepackage{lineno}
\linenumbers

\usepackage{longtable}

\usepackage{relsize}

\articletype{Research article}%
\received{XXXXXXXXX}
\revised{XXXXXXXXX}
\accepted{XXXXXXXXX}
\begin{document}

\lstdefinestyle{mystyle}{style=stata-editor, 
basicstyle=\footnotesize\ttfamily, 
numbers=left, 
numberstyle=\tiny\color{gray}, 
stepnumber=1, 
numbersep=5pt, 
backgroundcolor=\color{white}, 
showspaces=false, 
showstringspaces=false, 
showtabs=false, 
literate={`}{\textasciigrave}1,
frame=single, 
rulecolor=\color{black}, 
tabsize=1, 
captionpos=b, 
breaklines=true, 
extendedchars=true,
}

\lstset{style=mystyle}

\title{Performance of Cross-Validated Targeted Maximum Likelihood Estimation}

\author[1]{Matthew J. Smith*}

\author[2]{Rachael V. Phillips}

\author[1]{Camille Maringe$^{\pm}$}

\author[1,3]{Miguel Angel Luque-Fernandez$^{\pm}$}

\authormark{Smith \textsc{et al.}}

\address[1]{\orgdiv{Inequalities in Cancer Outcomes Network}, \orgname{London School of Hygiene and Tropical Medicine}, \orgaddress{\state{London}, \country{England, United Kingdom}}}

\address[2]{\orgdiv{Department of Biostatistics, School of Public Health}, \orgname{University of California at Berkeley}, \orgaddress{\state{California}, \country{United States of America}}}

\address[3]{\orgdiv{Department of Statistics and Operations Research}, \orgname{University of Granada}, \orgaddress{\state{Granada}, \country{Spain}}}

\corres{*Matthew J. Smith. \newline
\email{matt.smith@lshtm.ac.uk}}

\presentaddress{London School of Hygiene and Tropical Medicine, Keppel Street, London, England, United Kingdom, WC1E 7HT \\
\newline
$^{\pm}$ Senior authors contributed equally to this work.}

\abstract[Abstract]{
\textbf{Background} 
Advanced methods for causal inference, such as targeted maximum likelihood estimation (TMLE), require certain conditions for statistical inference. 
However, in situations where there is not differentiability due to data sparsity or near-positivity violations, the Donsker class condition is violated. 
In such situations, TMLE variance can suffer from inflation of the type I error and poor coverage, leading to conservative confidence intervals. 
Cross-validation of the TMLE algorithm (CVTMLE) has been suggested to improve on performance compared to TMLE in settings of positivity or Donsker class violations. 
We aim to investigate the performance of CVTMLE compared to TMLE in various settings. 
\newline
\textbf{Methods}
We utilised the data-generating mechanism as described in Leger et al. (2022) to run a Monte Carlo experiment under different Donsker class violations. 
Then, we evaluated the respective statistical performances of TMLE and CVTMLE with different super learner libraries, with and without regression tree methods. 
\newline
\textbf{Results}
We found that CVTMLE vastly improves confidence interval coverage without adversely affecting bias, particularly in settings with small sample sizes and near-positivity violations. Furthermore, incorporating regression trees using standard TMLE with ensemble super learner-based initial estimates increases bias and variance leading to invalid statistical inference. 
\newline
\textbf{Conclusions}
It has been shown that when using CVTMLE the Donsker class condition is no longer necessary to obtain valid statistical inference when using regression trees and under either data sparsity or near-positivity violations. We show through simulations that CVTMLE is much less sensitive to the choice of the super learner library and thereby provides better estimation and inference in cases where the super learner library uses more flexible candidates and is prone to overfitting.
\newline
}


\maketitle

\section{Introduction}\label{intro}

In public health research, it is often of interest to assess the causal relationship between an exposure or treatment and an outcome. Examples include the causal effect of immunotherapy on probability of survival after cancer diagnosis, the effect of smoking on rheumatoid arthritis, or the effect of childhood adversities on mental health later in life. Estimates of these relationships can be biased, such as spurious associations if there are factors that influence both the treatment and outcome variables. Randomised controlled trials (RCTs) minimise confounding due to randomisation of the individual to the treatment group. However, RCTs are not always feasible, such as for ethical reasons, or the randomisation process may fail. When causality cannot be guaranteed by design, such as in observational studies, causal inference methods must be used.\citep{Smith2022} \\

Methods used to estimate these causal effects can be broadly categorised into those that estimate the exposure model based on propensity scores, \cite{Austin_2011,Williamson_2011,Robins_2000,ROSENBAUM_1983} outcome model based on g-computation,\cite{Robins_1986,Snowden_2011,Vansteelandt_2011} or double robust methods, a combination of both exposure and outcome models.\cite{Bang_2005,Neugebauer_2005,vanderLaan2006} Double robust methods are so named because they are consistent estimators of the causal effect as long as at least one of the two models is correctly specified. Of the double robust methods, targeted maximum likelihood estimation (TMLE) has been shown to consistently provide the least biased estimate of the causal effect in comparison to other double robust methods such as inverse probability treatment weighting with regression adjustment (IPTW-RA) or augmented inverse probability treatment weighting (AIPTW).\cite{vanderLaan2006} The advantages of TMLE have been demonstrated theoretically, and in numerous simulation studies and applied analyses:\citep{vanderLaan2011TargetedLearning,Smith_2023} A plug-in estimator, TMLE respects the global limits of the statistical model (e.g., limiting the possible range of the targeted parameter). TMLE reduces bias through the use of ensemble and machine-learning algorithms and it has the minimum asymptotic variance in the class of semiparametric estimators. Statistical inference may be based on the efficient influence curve (IC) or bootstrap.\cite{vanderLaan2006,vanderLaan2011TargetedLearning,VanDerLaan2007SuperLearner,vanderLaan2018TargetedLearning, Cai2020a} The TMLE framework can be adapted for a wide range of causal effects, such as time-varying effects, dynamic treatment regimes, mediation analysis, amongst others. However, we focus only on point-treatment effects and the use of TMLE in estimating the average treatment effect (ATE). \\

The TMLE framework uses data adaptive ensemble machine learning algorithms for prediction of the outcome and treatment models, both considered nuisance models.\cite{Tsiatis2006} These nuisance models need to remain relatively simple to avoid overfitting, satisfy the Donsker class condition, and ensure valid inference. The Donsker class condition is a technical property that ensures asymptotic validity, such as the Gaussian process, for the true value of the parameter of interest. It can be seen as an extension of the Central Limit Theorem for functionals to compute Wald-type confidence intervals based on the IC and the functional Delta Method. Under the Donsker class condition, which requires data smoothness and differentiability, TMLE has desirable asymptotic properties such as consistency and normality.\cite{Zepeda-Tello2022} \\

Violation of the Donsker class condition are due to lack of differentiability, due to data sparsity (i.e., noncontinuous or step functions) or near-positivity violations, or when using aggressive machine learning algorithms such as the Random Forest. In situations where the Donsker class condition is violated, the variance can suffer from type I error inflation, leading to conservative confidence intervals and poor coverage.\cite{Zivich2021}  Cross-validated TMLE (CVTMLE) has been proposed to improve on performance in settings of Donsker class violations.\cite{LiHubbard2022} Cross-validation is a statistical learning technique widely used in regression and classification problems to avoid over-fitting and improve the asymptotic consistency and efficiency of estimations.\cite{Breiman_2017} \\


There are a couple of approaches to CVTMLE. One approach is based on Zheng \& van der Laan (2010)\cite{Zheng2010AsymptoticEstimation} who propose cross validating the entire TMLE algorithm and averaging all estimated treatment effects and their variances. More recently, Levy (2018) suggested that cross-validating the initial outcome model (which we denote as CVTMLE[Q]) would be sufficient for a more computationally efficient estimation of the target parameter, while retaining the theoretical properties of TMLE, particularly in cases where more complex machine learning algorithms are required.\cite{Levy2018AnCV-TMLE} \\


We aim to investigate the performance of CVTMLE[Q] compared to TMLE in settings with varying degrees of violation of the Donsker class condition. In Section 2, we describe TMLE and its cross-validated version. In Section 3, we outline the simulations of different settings likely violating the Donsker class condition. In Section 4 we report the respective performances of TMLE and CVTMLE[Q] when using different SuperLearner libraries. In Section 5, we reflect on the meaning of our results for practice and provide specific guidance.

\newpage

\section{Methods}
\subsection{Targeted Maximum Likelihood Estimation}

TMLE is a plug-in, semi-parametric, double-robust method that reduces the bias of an initial estimate by allowing for flexible estimation using nonparametric data-adaptive machine-learning methods to target an estimate closer to the true model specification.\cite{vanderLaan2006} Several tutorials for TMLE have been published along with a systematic review describing its applications.\cite{Smith2022,Smith_2023,Luque-Fernandez2018,Schuler2017,Gruber_2009} \\

TMLE is described in the \textit{Targeted Learning} book by Rose and van der Laan.\cite{vanderLaan2011TargetedLearning} We briefly outline the algorithmic steps when using TMLE for the ATE here. Given the data structure $O = (W,A,Y)$ observed on \textit{n} individual records, where \textbf{W} represents a set or vector of confounders, $A$ is a binary treatment or exposure mechanism, and $Y$ is the outcome, we suppose our target parameter is the average effect of treatment (ATE), across individuals. Using the potential outcomes framework, each individual has two potential outcomes: the outcome that would have been observed had the individual been exposed ($A = 1$) denoted as $Y(1)$, and the outcome that would have been observed had the individual not been exposed ($A = 0$) denoted as $Y(0)$. \\

\paragraph{Step 1: Predict the outcome}

TMLE fits the outcome model (i.e., $\mathrm{Q}^{0}(A,\mathbf{W}) = E(Y|A,\mathbf{W})$) using the observed values of the outcome, given observed treatment $A$ and covariates $\mathbf{W}$. To minimise model misspecification, an ensemble of machine-learning algorithms (i.e., Super Learner) is used to estimate $E(Y|A,\mathbf{W})$. Super Learner uses cross-validation to find the best-fitting combinations of models from a range of machine-learning algorithms to provide initial predictions of the outcome for each individual $n$ (i.e., $\mathrm{Q}^{0}_{n}(A_{i},\mathbf{W}_{i})$).\cite{vanderLaan2011TargetedLearning,VanDerLaan2007SuperLearner} \\

\paragraph{Step 2: Predict the treatment}

The Super Learner is also used to fit the propensity score model for the treatment (i.e., $ g(A,\mathbf{W}) = P(A = 1 \mid \mathbf{W})$) and predict treatment for each individual $n$ (i.e., $g(A = 1|\mathbf{W}_{i})$).\cite{vanderLaan2011TargetedLearning,VanDerLaan2007SuperLearner} \\

\paragraph{Step 3a: Calculate clever covariates}

Clever covariates (i.e., $H(A,W)$) are calculated using information from the observed treatment and predictions from the propensity score model. \\

$ H(1,\mathbf{W}) = \frac{A}{g(1,\mathbf{W})} $ \quad and \quad $ H(0,\mathbf{W}) = \frac{1-A}{g(0,\mathbf{W})} $ \\

\paragraph{Step 3b: Estimate the fluctuation parameter}

The fluctuation parameter ($\epsilon = \{\epsilon_{0},\epsilon_{1}\} $) is estimated through a maximum likelihood procedure: the observed outcome (Y) is regressed against the clever covariates with the logit of the initial prediction of $ \mathrm{Q}^{0}_{n}(A_{i},\mathbf{W}_{i}) $ as an offset. \\ \,


$ \mathrm{E}(\mathrm{Y}=1 \mid \mathrm{A}, \mathbf{W})(\varepsilon) \quad = \quad \frac{ { 1 } }{ 1 \,\, + \,\, \exp \left(-\log \left(\frac{\mathrm{Q}^{0}_{n}(A_{i},\mathbf{W}_{i})}{\left(1 \,\, - \,\, \mathrm{Q}^{0}_{n}(A_{i},\mathbf{W}_{i})\right)}\right) \,\, - \,\, \varepsilon_{0} \mathrm{H}(0, \mathbf{W}) \,\, - \,\, \varepsilon_{1} \mathrm{H}(1, \mathbf{W})\right) } $  \\ \\

When there is negligible remaining variability in $ Y - \mathrm{Q}^{0}_{n}(A_{i},\mathbf{W}_{i}) $, the fluctuation parameter will be estimated as close to 0. \\

\paragraph{Step 4: Update the initial predictions of the outcome}

The fluctuation parameter is used to update the initial outcome predictions from $\mathrm{Q}^{0}_{n}(A_{i},\mathbf{W}_{i})$ to $\mathrm{Q}^{1}_{n}(A_{i},\mathbf{W}_{i})$, optimising the bias-variance trade-off for the targeted parameter (average treatment effect [ATE]):\\

For any $A=a$: \quad $ \mathrm{Q}^{1}_{n}(A, \mathbf{W}_{i}) = expit\left(logit\left(Q^{0}_{n}(A, \mathbf{W}_{i})\right) + \frac{\varepsilon_{a}}{g(A, \mathbf{W})} \right) $ \\


\paragraph{Step 5: Estimate the target parameter}

Plug in the updated estimates of the predicted outcomes to the target parameter mapping for the ATE: \,

$$ \widehat{\operatorname{ATE}} = \frac{1}{\mathrm{n}} \sum_{\mathrm{i}=1}^{\mathrm{n}} \bigg( \mathrm{Q}^{1}_{n}\left(1, \mathrm{~\mathbf{W}}_{\mathrm{i}}\right)-{\mathrm{Q}}^{1}_{n}\left(0, \mathrm{~\mathbf{W}}_{\mathrm{i}}\right) \bigg) $$ \\

\paragraph{Step 6a: Estimate the efficient influence curve}

To calculate 95\% confidence intervals for the ATE, TMLE requires an estimate of the standard error for the ATE. The standard error is calculated based on the efficient influence curve (IC), which characterises the variability and represents the most efficient function.\cite{vanderLaan2011TargetedLearning,Zepeda-Tello2022,tlverse2023,Hampel1974} The efficient IC identifies how much influence a single data point has on the performance of TMLE in estimating the ATE, it is given by



$$ \widehat{\mathrm{IC}} = \left(\frac{\mathrm{A}}{g(1, \mathbf{W})}-\frac{1-\mathrm{A}}{g(0, \mathbf{W})}\right)\left(\mathrm{Y}-\mathrm{Q}^{1}(\mathrm{A}, \mathbf{W})\right) + \mathrm{Q}^{1}(1, \mathbf{W})-\mathrm{Q}^{1}(0, \mathbf{W})-\widehat{\mathrm{ATE}} $$ \\

The efficient IC combines information from the outcome model (Step 1 and 4), the propensity score model (Step 2), and the estimate of the target parameter (Step 5) to account for the variability in the estimator.  

\paragraph{Step 6b: Calculate standard error}

Then, the standard error ($\widehat{\sigma}_{\mathrm{ATE}}$) for the ATE is evaluated as: \\

$$ \widehat{\sigma}_{\mathrm{ATE}} = \sqrt{\frac{\widehat{{Var}}\left(\widehat{{IC}}_{\mathrm{ATE}}\right)}{\mathrm{n}}} $$ \\

where $\widehat{{Var}}\left(\widehat{{IC}}_{\mathrm{ATE}}\right)$ is the sample variance of the estimated IC.  \\

\paragraph{Step 6c: Calculate confidence intervals}

The 95\% confidence interval for the ATE is calculated as: \\

$$ \quad \quad  95\% \, CI \,\, = \,\, \widehat{ATE} \,\, \pm \,\, 1.96 \left( \widehat{\sigma}_{\mathrm{ATE}} \right)  $$ \\

\subsection{Cross Validated Targeted Maximum Likelihood Estimation}\label{CVTMLESL}

TMLE is a double-robust and efficient estimator but is susceptible to performance issues when the initial estimator of the outcome model is too adaptive. In other words, if the initial estimator of the outcome model is overfit, then there is negligible residual variation remaining for the targeting step.\cite{Li2022} Combining cross-validation with TMLE addresses this issue because training and validation are performed on indepedent sample subjects, which retains a realistic residual variation in the validation set. \\

There are several approaches to CVTMLE, each differ by what steps within TMLE are cross-validated.\cite{Zheng2010AsymptoticEstimation,Levy2018AnCV-TMLE} All approaches start with $K$ splits of the data. Each $k$ (with $k=1...K$) split defines each $k$ ($k=1...K$) fold, an indexing of the data into $k$ sets for algorithm training and validation. For a $K$-fold cross-validation scheme, the data is split evenly into $K$ subsets, the validation set for a given fold $k$ ($V_k$) is defined by the data in subset $k$, and the data not in subset $k$ is the training set for fold $k$ ($T_k$). Each subject is part of one validation set and $K-1$ training sets. \\

\begin{figure}[ht!]
    \centering
    \includegraphics[scale=1.1]{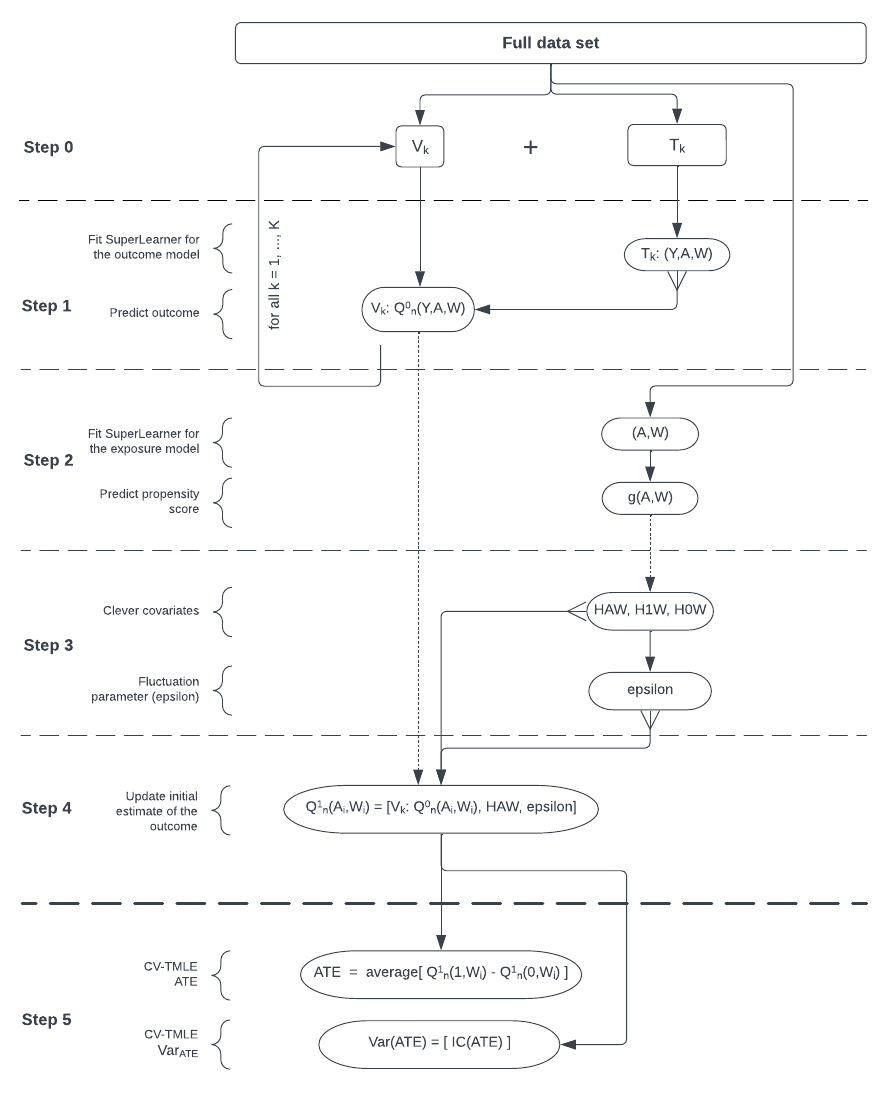}
    \caption{Process map of cross-validated targeted maximum likelihood estimation}
        \label{fig:CVTMLE}
    \end{figure}

We present one approach to CVTMLE that was proposed by Levy (2018),\cite{Levy2018AnCV-TMLE} which is adapted from the approach by Zheng \& van der Laan.\cite{Zheng2010AsymptoticEstimation} This approach makes use of cross-validation for estimating the outcome model only, and we denote this approach CVTMLE[Q]. The process for performing CVTMLE[Q] is illustrated in Figure \ref{fig:CVTMLE}. CVTMLE[Q] imposes that Step 1 of the TMLE algorithm described earlier is modified to accommodate K-fold cross-validation of the initial estimation of the outcome. For each cross-validation scheme, $k$, [$k=1...K$], estimate the outcome model (e.g., using the SuperLearner) using the training set, $Q_{T_k}^{0}(A,\textbf{W})$. From this initial model, the outcome is predicted for all observations within the corresponding validation set, $Q_{T_k}^{0}(a_{V_k},\textbf{w}_{V_k})$. This process is repeated for each cross-validation fold until each of the \textit{n} observations in the original data set has a predicted initial outcome $Q^{0}_{n}(A_{i},\textbf{W}_{i})$ for each individual $i$. The rest of the algorithm, steps 2-6c, proceeds as in the standard TMLE algorithm. Levy highlights that although predictions from the cross-validated sets are stacked, CVTMLE[Q] preserves the plug-in characteristic of the TMLE estimator and performs well asymptotically.\cite{Levy2018AnCV-TMLE}

\newpage ~\\
\newpage 

\section{Simulations}\label{simulations}

\subsection{Setting}\label{setting}

To evaluate the performance of TMLE and CVTMLE under near-positivity violations, we perform a Monte Carlo simulation experiment in which we varied the likely severity of the violation of the Donsker class condition. There are different situations in which models may not satisfy the Donsker class condition: i) data sparsity or small sample size, ii) near-positivity violations, and iii) the use of highly data-adaptive machine learning algorithms (e.g., tree-based algorithms, such as random forests), all leading to non-differentiability of the influence function. Table \ref{tab:DCCviolation} expands on these different scenarios leading to violation of the Donsker class condition and how the simulations were specified to force the violation of the Donsker class condition.

\begin{table}[ht!]
\centering
\caption{Settings where the Donsker class condition is likely violated and how these were reproduced in simulations.}
\label{tab:DCCviolation}
\begin{tabular}{llll}
\hline
\\
\textbf{Setting} &
  \textbf{Description} &
  \textbf{\begin{tabular}[t]{@{}l@{}} Impact on Donsker \\ class condition \end{tabular}} &
  \textbf{Simulation} 
    \\
    \\ \hline
  
  & \\
    & \\
  
\begin{tabular}[t]{@{}l@{}} Sample size\end{tabular} &
  \begin{tabular}[t]{@{}l@{}} Small sample size requires \\ a greater number of \\ folds to be used within \\ CVTMLE to allow a \\ large enough training set. \\ \\ \end{tabular} &
  \begin{tabular}[t]{@{}l@{}} Donsker class condition is \\ based on asymptotic theory, \\ which assumes that the \\ sample size goes to infinity. \\ \\ Small sample size can lead \\ to random noise dominating \\ the signal that machine \\ learning algorithms are \\ attempting to model. \end{tabular} &
  \begin{tabular}[t]{@{}l@{}} (i) Large sample size  \\ (n=1000) that does not \\ require an increase in the \\ number of folds (default of \\ 10 folds is used). \\ \\ (ii) Small sample size (n=200) \\ that requires an increase in \\ the number of folds but is \\ kept at the default of 10 folds.\end{tabular} \\ 
  
  & \\ \hdashline
  & \\
  
\begin{tabular}[t]{@{}l@{}}Near-positivity \\ violation\end{tabular} &
  \begin{tabular}[t]{@{}l@{}} There are groups of \\ individuals with near-zero \\ probability to be treated \\ or untreated, which leads \\ to gaps in the data \\ with unobserved or \\ impossible combinations of \\ the exposure/outcome. \end{tabular} &
  \begin{tabular}[t]{@{}l@{}} Near-positivity violations \\ can introduce abrupt \\ changes, discontinuities, or  \\ irregularities in the \\ empirical process (i.e., \\ estimation of the influence \\ function), disrupting its \\ smooth convergence.\end{tabular} &
  \begin{tabular}[t]{@{}l@{}} (i) High prevalence of \\ A (i.e., P{[}A=1{]}=0.8) \\ created in exposure model.\\ \\ (ii) Extrapolation issue \\ created by interaction in the \\ outcome model between \\ treatment and rare covariate.\end{tabular} \\ 
  
  & \\ \hdashline
  & \\
  
\begin{tabular}[t]{@{}l@{}}Complex machine \\ learning algorithm\end{tabular} &
  \begin{tabular}[t]{@{}l@{}}Machine learning methods, \\ such as tree-based algorithms \\ (e.g., random forests) used \\ in the SuperLearner for the \\ outcome and propensity \\ score models\end{tabular} &
  \begin{tabular}[t]{@{}l@{}} Tree-based methods \\ are highly data-adaptive \\ and have a tendency to \\ overfit the data, \\ especially in smaller \\ sample sizes. \end{tabular} &
  \begin{tabular}[t]{@{}l@{}} Using random forests \\ with and without cross- \\ validation of TMLE to see \\ the impact of \\ cross-validation on \\ variance stabilisation.\end{tabular} \\ 
  
  & \\ \hdashline
  & \\   
  
\begin{tabular}[t]{@{}l@{}}Non-differentiability of the \\ Influence function (IF)\end{tabular} &
  \begin{tabular}[t]{@{}l@{}} Influence function (a.k.a. \\influence curve) must be \\ continuous at every point \\ in its domain, but fails to be \\ differentiable at a bend, cusp, \\ or vertical tangent.\end{tabular} &
  \begin{tabular}[t]{@{}l@{}} IF is derived based on limiting \\ behaviour of the estimator. \\ When Donsker class condition \\ is violated, the empirical \\ process does not converge to \\ a smooth limiting distribution. \end{tabular} &
  \begin{tabular}[t]{@{}l@{}} Combination of small sample \\ size, near-positivity violation, \\ and complex machine learning \\ algorithms used to estimate \\ the target parameter. \end{tabular} \\
  & \\ \hline
  
\end{tabular}
\end{table}

\subsection{Data generating mechanisms}\label{DGM}

We simulated scenarios of near-positivity violations using data-generating mechanisms described in Leger \textit{et al} (2022).\cite{Leger2022}  \\

First, we generated a vector of covariates $\mathbf{W} = {W_{1},W_{2},W_{3},W_{4},W_{5},W_{6},W_{7},W_{8}}$, including six binary covariates following Bernoulli distributions with probabilities 0.1 for $W_{1}$, 0.4 for $W_{2}$, 0.7 for $W_{4}$, 0.5 for $W_{5}$, 0.3 for $W_{7}$, 0.8 for $W_{8}$, and two continuous covariates, $W_{3}$ and $W_{6}$ following a Gaussian distribution with mean 0 and standard deviation 1. \\

The exposure $A$ was generated according to a Bernoulli distribution with probability obtained from a logistic regression model, using a logit link function, with the following linear predictor: $\alpha_{0} + \alpha_{1}W_{1} + \alpha_{2}W_{2} + \alpha_{4}W_{4} + \alpha_{6}W_{6} + \alpha_{7}W_{7} + \alpha_{8}W_{8}$. Where $\alpha_{0}$ was set to -0.45 or 1.05 to simulate prevalence of exposed patients at 50\% or 80\%, respectively. $\alpha_{1}$, the coefficient for $W_{1}$ was set to $log(5)$ to impose a near-positivity violation particularly given that $W_{1}$ is generated with 10\% prevalence. The rest of the coefficients, $\alpha_{2},\alpha_{4},\alpha_{6},\alpha_{7},\alpha_{8}$, were set to log(1.5). \\

Near-positivity violation was determined from the values of the propensity scores (appendix table \ref{tab:PStable}) that were greater than the cut-off for truncation at 0.975. With 80\% prevalence of the exposure there was, on average, 2.2 and 10.8 propensity scores that exceeded 0.975 for samples of 200 and 1000, respectively. With 50\% prevalence of the exposure there were, on average, no propensity scores larger than the cut-off for truncation. \\



The outcome was generated from a Bernoulli distribution with probability obtained from a logistic regression model, using a logit link function, with the following linear predictor: $-0.8 + \beta_{A}A + \beta_{1}W_{1} + \beta_{2}W_{2} + \beta_{3}W_{3} + \beta_{4}W_{4} + \beta_{5}W_{5} + \beta_{6}W_{6} + \beta_{7}A \times W_{1}$. $\beta_{A}$, the coefficient for the exposure was set to $log(1.75)$. The interaction term $A \times W_{1}$ is included with coefficient $\beta_{7}$ set at 0 or 2 for the absence or presence of an extrapolation issue, respectively. The rest of the coefficients were set to $log(1.5)$. The distribution for the probability of the outcome is shown in appendix figure \ref{fig:ProbY}. \\

We simulated datasets of sample sizes $n_{obs} = \{200, 1000\}$ representing small and large sample sizes, respectively, based on Leger \textit{et al}.\cite{Leger2022} We chose a large enough sample of repetitions ($n_{reps} = 1000$) such that we obtained a small enough Monte Carlo standard error without unfeasible computational time even for $n_{obs} = 200$. The formula for the 95\% confidence interval around the mean estimate is:\cite{Tang2005ATrial} \,
        
        $$ p \pm 1.96 * \sqrt{\frac{p(1-p)}{n_{reps}}} $$  \,
        
Substituting $p$ with the nominal coverage probability, 0.95 or 95\%, the estimated coverage should fall between 93.6\% and 96.4\%. 

\newpage ~\,
\newpage

\subsection{Estimand, Methods, and Performance Measures}\label{performance}

The estimand of interest was the average treatment effect (ATE) estimated by the difference in risks of the outcome between exposed and unexposed, $\hat{\theta} = \hat{\pi_{1}} - \hat{\pi_{0}}$. The true risk difference ($\theta$) of the outcome between the exposed ($\pi_{1}$) and unexposed ($\pi_{0}$) was generated from the exposure and outcome models defined above, using data from the respective repetition, and calculated by averaging the repetition-specific true risk differences ($\theta_{i}$). wo different estimation methods: TMLE and CVTMLE[Q]. Both estimation methods were used, by default, with the following algorithms within the SuperLearner: i) stepwise selection, ii) generalized linear modelling (glm), iii) a glm variant that included second order polynomials and two-by-two interactions of the main terms included in the models. We also included additional algorithms within the SuperLearner such as Lasso (\textit{glmnet} R package), Random Forest (\textit{randomForest} R package), and Generalised Additive Models [all of which referred to as 'RF']. Therefore, the performances of four methods were contrasted: TMLE, CVTMLE[Q], TMLE-RF, and CVTMLE[Q]-RF. All simulated variables (i.e., $W_{1}, W_{2}, W_{3}, W_{4}, W_{5}, W_{6}, W_{7}, W_{8} $) were included \textit{a priori} for all estimation methods.\\

We assessed the performance of each method using measures of confidence interval coverage, relative error, and relative bias.\cite{Morris2019} The confidence interval coverage is the proportion of confidence intervals estimated around each repetition-specific estimate $\hat{\theta}$ (i.e., $\hat{lower}^{\hat{\theta}}_{1-\alpha},\hat{upper}^{\hat{\theta}}_{1-\alpha}$) that include the true ATE ($\theta$). It is calculated as: 

$$ \text{Coverage} \quad = \quad \operatorname{Pr}\left(\hat{\theta}_{\text {low}} \leq \theta \leq \hat{\theta}_{\mathrm{upp}}\right) \quad  =  \quad \frac{1}{n_{\mathrm{reps}}} \sum_{i=1}^{n_{\mathrm{reps}}} 1\left(\hat{\theta}_{\text {low}, i} \leq \theta \leq \hat{\theta}_{\mathrm{upp},i}\right) $$ \\ \,

Ideal confidence interval coverage is near $1-\alpha$, where $\alpha$ is usually chosen as 0.05. To reach nominal coverage, we expect that $95\%$-confidence intervals would cover the true ATE in $95\%$ of the repetitions. \\

fine the relative error, we need to define (i) the model-based standard error (ModSE): the square-root of the average of the repetition-specific variances, and (ii) the empirical standard error (EmpSE): the standard error of the target parameters across the repetitions. The relative error is the relative percentage error in ModSE in its estimation of EmpSE, such that if ModSE is an appropriate estimate of the variance, the relative error will be close to 0 while if ModSE systematically fails to estimate the EmpSE, it represents a bias in the estimation of the model-based standard error. The relative error is calculated as:

$$ \text{Relative error} \quad = \quad 100\left(\frac{\mathrm{ModSE}}{\mathrm{EmpSE}}-1\right) \quad = \quad 100\left(\frac{\sqrt{\frac{1}{n_{\mathrm{reps}}} \sum_{i=1}^{n_{\mathrm{reps}}} \widehat{\operatorname{Var}}\left(\hat{\theta}_{i}\right)}}{\sqrt{\frac{1}{n_{\mathrm{reps}}-1} \sum_{i=1}^{n_{\mathrm{reps}}}\left(\hat{\theta}_{i}- E[\hat{\theta}]\right)^{2}}}-1\right) $$ \\ \,

\noindent where $\hat{\theta}_{i}$ is the estimate of the ATE $\theta$ from the $i^{th}$ repetition, and $E[\hat{\theta}]$ is the mean of $\hat{\theta}_{i}$ across repetitions. \\

The relative bias is the relative difference between the estimated ATE, $E[\hat{\theta}]$, and the true value of the ATE, ($\theta$), and is calculated as: 

$$ \text{Relative Bias} \quad = \quad \frac{E[\hat{\theta}]-\theta}{\theta} \quad = \quad \frac{\frac{1}{n_{\mathrm{reps}}}\sum_{i=1}^{n_{\operatorname{reps}}} (\hat{\theta}_{i}-\theta_{i})}{\frac{1}{n_{reps}}\sum_{i=1}^{n_{reps}}\theta_{i}} $$ \\  \,

All analysis were performed in Stata statistical software (StataCorp, 2020. StataCorp LLC, College Station, TX). The Stata code to run the simulations is available at: \hyperlink{https://github.com/mattyjsmith/CVTMLE}{https://github.com/mattyjsmith/CVTMLE} We used the \textit{eltmle} command to perform all methods.\cite{LuqueFernandez2021} Recent updates include the functionality to assess positivity violations via covariate balance tables. The command has been updated to perform cross-validated TMLE (for the outcome model only, CVTMLE[Q]) but is not yet publicly available and an update is in preparation.

\newpage 

\section{Results}\label{Results}

We report the performance measures for all simulated scenarios in Figures \ref{fig:Combined}, and interpret the coverage, relative bias, and relative error in turn in the following subsections. 

\subsection{Coverage}

Based on $n_{reps} = 1000$ repetitions, an optimal coverage is within the range of 93.6\% to 96.4\%, as indicated by the horizontal dashed lines (Figure \ref{fig:Combined}, row 1). TMLE has optimal coverage for large sample sizes but only when there is 50\% prevalence of the exposure in small sample sizes (Figure \ref{fig:Combined}, row 1, 50\% prevalence). With 80\% prevalence of the exposure (Figure \ref{fig:Combined}, row 1, 80\% prevalence), there is slight undercoverage in the absence of an extrapolation issue, but the magnitude of the undercoverage increases with high extrapolation issue. CVTMLE[Q] provides the most consistent coverage regardless the sample size, prevalence of the exposure, or extent of the extrapolation issue (Figure \ref{fig:Combined}, row 1). Only where there is small sample size and high extrapolation issue, and with 80\% prevalence of the exposure, CVTMLE[Q] provides a slight undercoverage. \\



When using complex tree-based algorithms within the SuperLearner, the use of cross-validation improves the coverage rate in comparison to standard TMLE, which shows undercoverage. CVTMLE[Q]-RF has optimal coverage with 50\% prevalence of the exposure, regardless of the presence of extrapolation issue or sample size (Figure \ref{fig:Combined}, row 1, 50\% prevalence). However, there is undercoverage with 80\% prevalence of the exposure and high extrapolation issue or small sample size, or both (Figure \ref{fig:Combined}, row 1, 80\% prevalence). 


\subsection{Relative bias}

There are no differences in relative bias performance between TMLE and TMLE[Q], and TMLE-RF and TMLE-RF[Q]. All methods show negligible bias when there is large sample size and no extrapolation issue (Figure \ref{fig:Combined}, row 2). With 50\% prevalence of the exposure, in large sample sizes, there is approximately 5\% relative bias for all methods in the presence of a high extrapolation issue (Figure \ref{fig:Combined}, row 2, 50\% prevalence). In addition, for all methods, the relative bias is approximately 10\% when there is 80\% prevalence of the exposure (Figure \ref{fig:Combined}, row 2, 80\% prevalence). The relative bias increases in the presence of an extrapolation issue, with small sample size, or both, and similar patterns are observed regardless of prevalence, with larger amplitude for settings with 80\% prevalence of the exposure. Relative bias increases when using additional tree-based algorithms in the SuperLearner.


\subsection{Relative error}

\begin{figure}[ht!]
        \centering
        \includegraphics[width=0.8\linewidth]{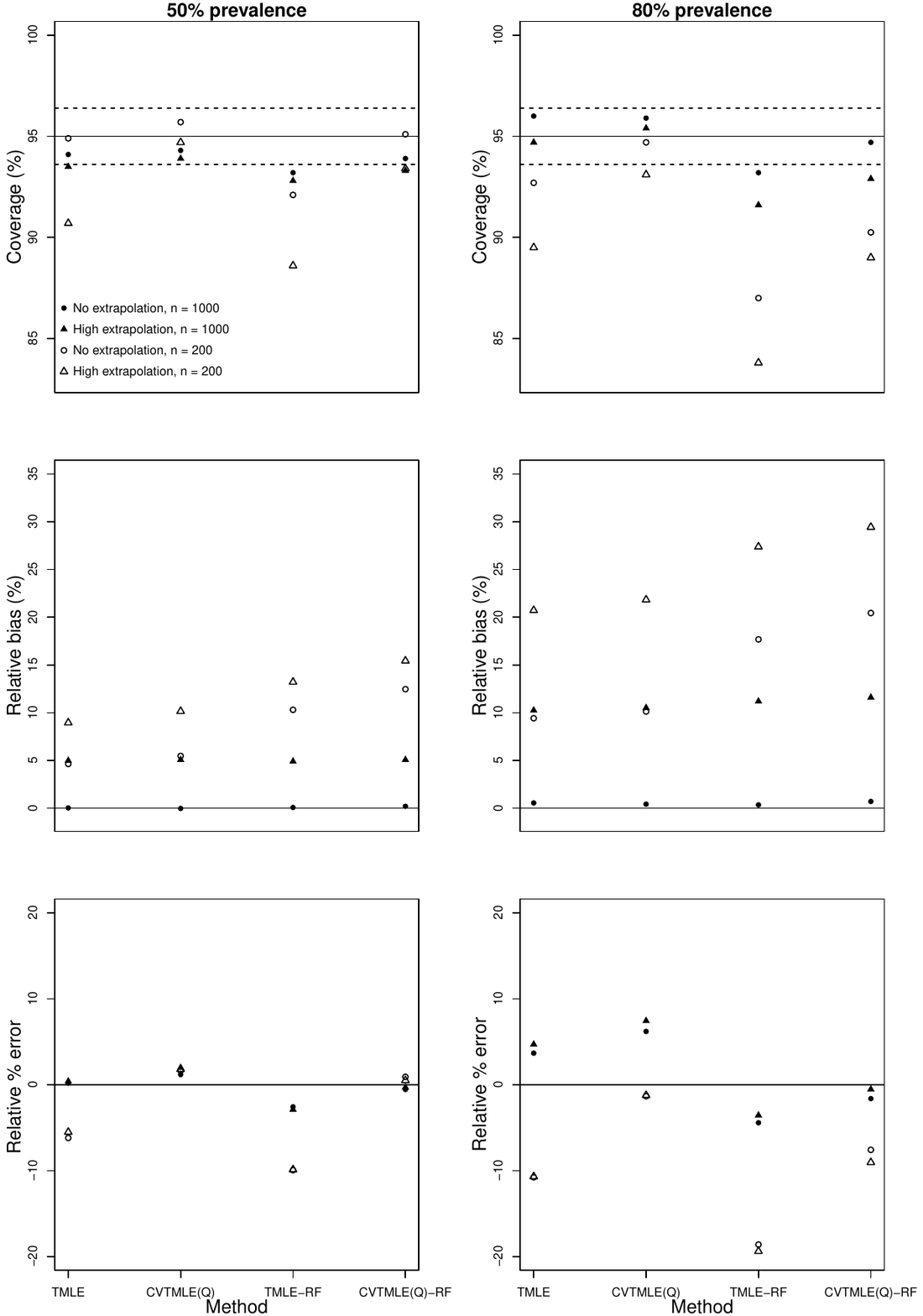}
        \caption{Coverage, relative bias, and relative percentage error (y-axes) of each method (x-axis) according to different sample sizes and the presence/absence of extrapolation issue, stratified by prevalence of the exposure (50\% or 80\%). Optimal coverage is 95\% (range: 93.6\% - 96.4\%) shown by solid and dashed horizontal lines.Cross-validation using 5 folds. }
        \label{fig:Combined}
    \end{figure}

The relative percentage error in ModSE is shown in Figure \ref{fig:Combined} (row 3). All methods exhibit comparable estimates of the EmpSE within each scenario (results not shown), with a slight reduction in EmpSE when using cross-validation. The difference in relative error between all of the methods is thus due to differences in ModSE. Regardless the prevalence of the exposure, CVTMLE[Q] and CVTMLE[Q]-RF consistently show the smallest relative percentage error in ModSE (Figure \ref{fig:Combined}, row 3). In general, scenarios with small sample sizes increase the magnitude of the relative error but the presence of extrapolation issue does not alter the relative percentage error.

\newpage ~\\
\newpage

\section{Discussion}\label{Discussion}

We found that combining targeted maximum likelihood estimation with cross-validation (CVTMLE) improves coverage without adversely affecting bias, particularly in settings of small sample sizes and near-positivity violations. In terms of bias and coverage, TMLE performs as well as CVTMLE in large sample sizes but suffers when the Donsker class condition is in question with undercoverage in cases of small sample sizes with extrapolation issues (i.e., data sparsity forced with an interaction in the outcome model between treatment and a rare covariate), or unbalanced prevalence of the exposure. \\

It has been advocated that researchers should use a richly specified library of machine learning algorithms within the SuperLearner to maximise the performance of the estimation approach.\cite{Naimi_2021} Previous research suggests that tree-based methods, such as random forests, should be used with care because they tend to overfit the data.\cite{Balzer2021b} In concordance, we found that the use of random forests led to a severe undercoverage when used with TMLE in settings likely violating the Donsker class condition. If tree-based methods must be utilised in the estimation step (i.e., due to the presence of heterogeneous treatment effects),\cite{Jawadekar_2023} we advocate for the use of cross-validation of the outcome model (CVTMLE[Q]) to optimise the estimation of the standard error and retrieve appropriate coverage.\\ 

\begin{figure}[ht!]
    \centering
    \includegraphics[width=1.0\linewidth]{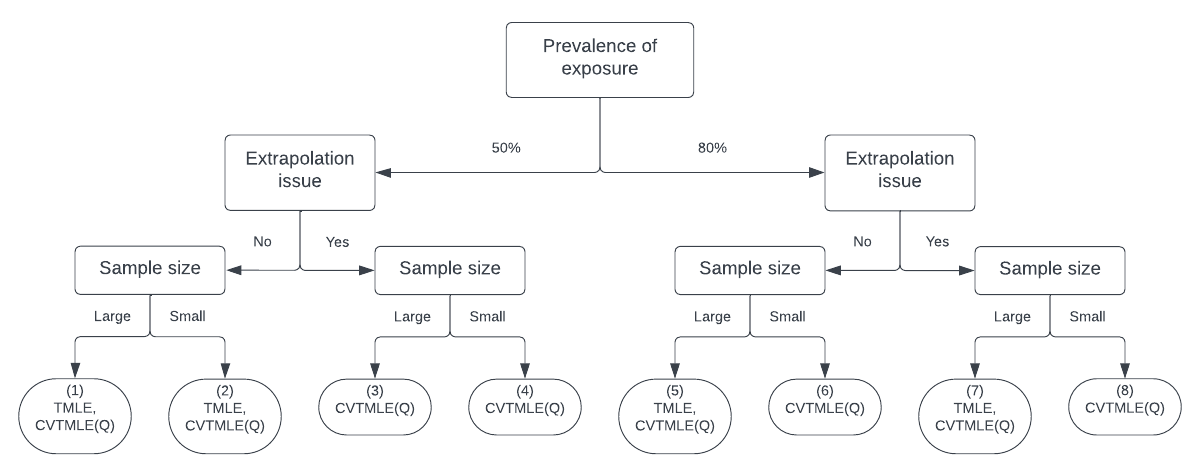}
    \caption{Decision tree for the appropriate choice of method given the scenarios (i.e., near-positivity violation, sample size) that can cause the lack of differentiability of the influence curve and potentially violate the Donsker class condition.}
    \label{fig:DecisionTree}
\end{figure}

As shown in this simulation study, the choice of the method to use is dependent on whether the data exhibits characteristics that could lead to violation of the Donsker class condition. We provide a decision tree to guide the choice of estimation method in applied settings depending on the prevalence of the exposure, the finite sample size and the presence of potential extrapolation and/or near positivity violations due to data sparsity (Figure \ref{fig:DecisionTree}). For example, in Branch (1) where there is 50\% prevalence of the exposure, no extrapolation issue, and large sample size, our results suggest that either of TMLE or CVTMLE[Q] could be chosen to obtain an unbiased estimate of the average treatment effect with optimal coverage. CVTMLE[Q] is a suitable choice for any of the branches and is often the only appropriate choice of these methods, particularly in settings with near-positivity violation and small sample sizes (such as in Branches 4, 6, and 8). However, cross-validation is computationally intensive and if there are other methods (e.g., standard TMLE) that would perform the analysis to a similar degree of accuracy, then these other methods could be considered. Such instances occur in Branches 2, 5, and 7, where TMLE is a suitable alternative to CVTMLE[Q] because it is as similarly least biased and within the optimal coverage range. 

We generated data with simple outcome and exposure models to focus on, and highlight, the improvements in coverage when using cross-validation with TMLE. Naimi \textit{et al} (2021) compared the performance of TMLE between simple and complex models.\cite{Naimi_2021} In our simulation study, we did not include complex terms other than an interaction between the exposure and a variable causing the near-positivity violation. Further studies are needed to explore the performance of these two methods in the context of data generated by complex models and heterogeneous treatment effects (i.e., inclusion of additional interactions, non-linear, and time-dependent effects). We speculate that methods employing additional algorithms (e.g., random forests) might perform better in terms of bias and, indirectly, coverage. Moreover, we considered only binary variables for the outcome and exposure. The performance of these methods in settings with a continuous exposure or outcome require further exploration: we speculate that the trends and patterns observed in this simulation study are generalisable to continuous outcomes and exposures but this requires further research to confirm. Further research could investigate a comparison of Zheng \& van der Laan's approach\cite{vanderLaan2011TargetedLearning} to Levy's approach\cite{Levy2018AnCV-TMLE} in settings with simple and complex outcome and exposure models. \\

An alternative approach to cross-validation is to use cross-fitting procedures. Similar to cross-validation, doubly robust cross-fit estimators have been developed to reduce overfitting and impose less restrictive complexity conditions on the machine learning algorithms used to estimate nuisance functions.\cite{Zivich2021,Chernozhukov2018Double/debiasedParameters,Newey_2017} There is a subtle difference in the terminology, sample-splitting is the procedure of cross-validation and cross-fitting, but they differ in their purpose. Sample splitting in cross-validation is used for model selection and validation of the nuisance parameters (i.e., the outcome and exposure models), whereas sample splitting in cross-fitting is used within TMLE (i.e., CVTMLE) to reduce bias in the estimation of the causal parameter by separating the estimation of nuisance parameters from the estimation of the target parameter.\cite{vanderLaanDML2019} \\


Previous research has shown smaller sample sizes require an increase in the number of folds when performing the Super Learner.\cite{Phillips2022PracticalLearner} This is to allow a sufficiently large training set to train the nuisance models. We did not alter the default setting of 10 folds used within the Super Learner but the benefit of correctly specifying the number of required folds for cross-validation within the Super Learner and the cross-validation of TMLE is an area of ongoing research. We contrasted 5 and 10 fold cross-validation schemes and did not notice differences in performances for CVTMLE[Q] and CVTMLE[Q]-RF. \\

While TMLE is available in several software,\cite{Pang2016a,zepid}, to our knowledge the functionality to cross-validate TMLE is limited to only Stata (\textit{eltmle}\cite{LuqueFernandez2021}) and R (\textit{tmle}\cite{GruberTMLE2012}, textit{tmle3}\cite{coyle2021tmle3-rpkg}). Importantly, TMLE R software defaults to CVTMLE[Q]. Other packages exist that can be adapted to cross-validate TMLE, such as Origami\cite{coyle2018origami} for TMLE3\cite{coyle2021tmle3-rpkg} in R, but tutorials are sparse. We used the \textit{eltmle} command in Stata where CVTMLE is not yet publicly available but is in preparation.\cite{LuqueFernandez2021} \\

This study was limited to only one estimator, but other double-robust estimators exist, such as augmented inverse probability of treatment weighting (AIPTW). We considered only TMLE-based methods because (i) of their better stability, and (ii) we aimed to specifically investigate the undercoverage of TMLE.\cite{Luque-Fernandez2018} CVTMLE helps to make the estimator consistent in larger samples, however performance issues may still occur for finite samples.\cite{Li2022} For example, if the data violates the positivity assumption (i.e., the probability of being exposed, or unexposed, is too close to 0 or 1), which is more likely in smaller samples, then instability of the the inverse weighting may occur in the targeting step. A simplistic approach is to truncate the propensity score at 0.975 and 0.025. However, collaborative-TMLE (C-TMLE) is another viable option:\cite{vanderLaan2011TargetedLearning,Stitelman2010,Benkeser_2020} C-TMLE adaptively estimates the propensity score based on the outcome regression and mitigates practical positivity violations.\cite{Balzer2021b} C-TMLE has been recently developed that performs a model selection in estimating the propensity score model, which prevents the targeting step from introducing instability into the estimator of the outcome model. In this study we focused on the comparison of TMLE and CVTMLE, further studies are needed to compare these other methods. \\

We observed that TMLE produces an underestimate of the coverage in settings with small sample sizes, presence of extrapolation issue, or imbalances prevalence of the exposure; however, combining cross-validation with TMLE allows a consistent and reliable estimate of the coverage. The analysis of high dimensional data is an increasingly common activity for applied researchers, which often requires handling complex relationships between variables, and is likely to incur many of the data-generating mechanisms employed in this simulation study. The implications of these findings suggest that it is not only important to check all necessary distributions (e.g., overlap plots) before estimating the effect of interest but that applied researchers should be cautious when choosing the appropriate method to analyse high-dimensional data and strongly consider using cross-validation, or similar, techniques to avoid issues with undercoverage (a consequence of high type I error) that occur in standard TMLE. 

\subsection{Conclusion}

In conclusion, our simulation study reveals the substantial benefits of incorporating targeted maximum likelihood estimation with cross-validation in addressing coverage issues, particularly for small sample sizes and near-positivity violations. Notably, the cross-validation of the outcome model (CVTMLE[Q]) consistently yielded optimal coverage estimates. Our results underscore the importance of cross-validation techniques, especially in the analysis of high-dimensional data, cautioning researchers to consider cross-validation to mitigate issues of undercoverage whenever TMLE is being used.

\newpage
\section*{Funding}
This work was supported by the Medical Research Council [grant number MR/W021021/1] and MCIN/AEI/
10.13039/501100011033. A CC BY or equivalent licence is applied to the Author Accepted Manuscript (AAM) arising from this submission, in accordance with the grant’s open access conditions. Camille Maringe is supported by a Cancer Research UK Population Research Committee Programme Award (C7923/A29018).

\section*{Authors contributions}
The article arose from the motivation to understand how cross-validated targeted maximum likelihood estimation performs in the presence of positivity violations. All authors developed the concept and MJS wrote the first draft of the article. MJS, RVP, MALF and CM revised the manuscript. All authors read and approved the final version of the manuscript. MJS is the guarantor of the article.

\section*{Acknowledgements}
The motivation and some parts of the manuscript come from MALF's work in a visiting academic position in the Division of Biostatistics at the Berkeley School of Public Health in 2019.
For the purpose of open access, the authors have applied a Creative Commons Attribution (CC BY) license to any Author Accepted Manuscript version arising.

\newpage

\bibliography{bibliography}

\newpage

\appendix

\section{Tables and figures}

\begin{table}[ht!]
\centering
\caption{Summary statistics of propensity scores by sample size, prevalence of the exposure "P(A=1)", and exposure group. The mean of "n \textgreater 0.975" is the average number of propensity scores, across the 1000 samples, that exceed the truncation value of 0.975. The range is the minimum and maximum number of propensity scores, across the 1000 samples, that exceed the truncation value of 0.975.}
\label{tab:PStable}
\begin{tabular}{ccclcccccc}
\textbf{} & \textbf{} & \textbf{} &  & \multicolumn{3}{c}{\textbf{Propensity scores}} & \textbf{} & \multicolumn{2}{c}{\textbf{n \textgreater 0.975}} \\ \cline{5-7} \cline{9-10} 
\textbf{}             & \textbf{}             & \textbf{}  &  & \textbf{Min} & \textbf{Mean} & \textbf{Max} & \textbf{} & \textbf{Mean} & \textbf{Range} \\ \cline{5-7} \cline{9-10} 
\textbf{Sample size}  & \textbf{P(A=1)}       & \textbf{A} &  & \textbf{}    & \textbf{}     & \textbf{}    & \textbf{} & \textbf{}     & \textbf{}      \\ \cline{1-3}
\multirow{5}{*}{200}  & \multirow{2}{*}{50\%} & 1          &  & 0.222        & 0.554         & 0.921        &           & 0.0           & (0, 0)         \\
                      &                       & 0          &  & 0.178        & 0.452         & 0.845        &           & 0.0           & (0, 0)         \\
                      &                       &            &  &              &               &              &           &               &                \\
                      & \multirow{2}{*}{80\%} & 1          &  & 0.513        & 0.812         & 0.981        &           & 2.2           & (0, 9)         \\
                      &                       & 0          &  & 0.510        & 0.752         & 0.931        &           & 0.0           & (0, 2)         \\
                      &                       &            &  &              &               &              &           &               &                \\
\multirow{5}{*}{1000} & \multirow{2}{*}{50\%} & 1          &  & 0.182        & 0.554         & 0.945        &           & 0.0           & (0, 1)         \\
                      &                       & 0          &  & 0.143        & 0.453         & 0.902        &           & 0.0           & (0, 0)         \\
                      &                       &            &  &              &               &              &           &               &                \\
                      & \multirow{2}{*}{80\%} & 1          &  & 0.454        & 0.813         & 0.987        &           & 10.8          & (2, 22)        \\
                      &                       & 0          &  & 0.440        & 0.753         & 0.963        &           & 0.2           & (0, 3)         \\ \cline{1-3} \cline{5-10} 
\end{tabular}
\end{table}

\begin{figure}[ht!]
    \centering
    \includegraphics[width=1.0\linewidth]{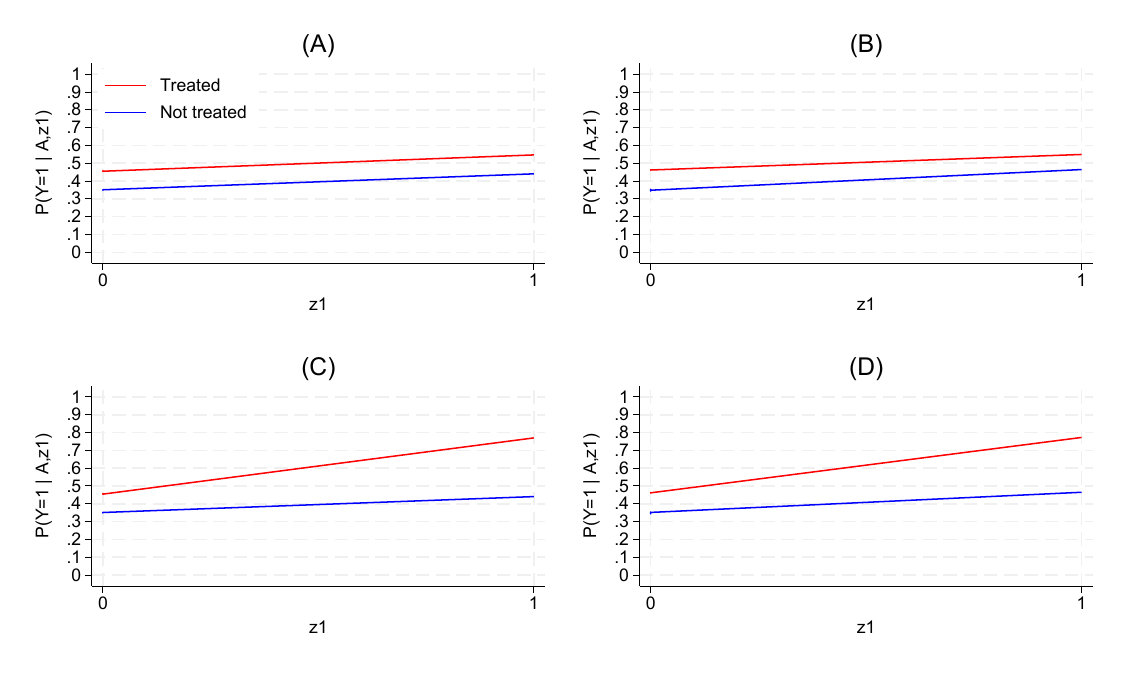}
    \caption{Probability of the outcome given the exposure and Z1 (variable creating near-positivity violations), stratified by prevalence of the exposure (i.e., 50\% or 80\%) and presence of extrapolation issue (i.e., none or high). (A) is 50\% prevalence of the exposure with no extrapolation issue. (B) is 80\% prevalence of the exposure with no extrapolation issue. (C) is 50\% prevalence of the exposure with an extrapolation issue. (D) is 80\% prevalence of the exposure with an extrapolation issue}
    \label{fig:ProbY}
\end{figure}

The probability of the outcome for 50\% prevalence is shown in Figure \ref{fig:ProbY}(A) and \ref{fig:ProbY}(C), and for 80\% prevalence is shown in Figure \ref{fig:ProbY}(B) and \ref{fig:ProbY}(D). A high extrapolation issue, created by an interaction between the exposure $A$ and $W_{1}$ in the outcome model, is shown in \ref{fig:ProbY}(C) and \ref{fig:ProbY}(D) and leads to non-parallel lines for the probabilities of the outcome by treatment group. There was no extrapolation issue generated in scenarios depicted in Figures \ref{fig:ProbY}(A) and \ref{fig:ProbY}(B). \\

\end{document}